\documentclass{emulateapj}

\shorttitle{Chandra Observations of SDSS~J1004+4112}
\shortauthors{Ota et al.}
\begin{document}

\title{Chandra Observations of SDSS~J1004+4112: Constraints on the
  Lensing Cluster and Anomalous X-Ray Flux Ratios of the Quadruply
  Imaged Quasar}

\author{
Naomi Ota,\altaffilmark{1}  
Naohisa Inada,\altaffilmark{2} 
Masamune Oguri,\altaffilmark{3} 
Kazuhisa Mitsuda,\altaffilmark{4} \\
Gordon T. Richards,\altaffilmark{5} 
Yasushi Suto,\altaffilmark{6} 
W. N. Brandt,\altaffilmark{7}
Francisco J. Castander,\altaffilmark{8} \\
Ryuichi Fujimoto,\altaffilmark{4}  
Patrick B. Hall,\altaffilmark{9}
Charles R. Keeton,\altaffilmark{10} 
Robert C. Nichol,\altaffilmark{11} \\
Donald P. Schneider,\altaffilmark{7} 
Daniel E. Eisenstein,\altaffilmark{12} 
Joshua A. Frieman,\altaffilmark{13} 
Edwin L. Turner,\altaffilmark{3}\\
Takeo Minezaki,\altaffilmark{2}
and 
Yuzuru Yoshii\altaffilmark{2,14}
}

\altaffiltext{1}{Cosmic Radiation Laboratory, RIKEN (The Institute of
  Physical and Chemical Research), 2-1 Hirosawa, Wako, Saitama
  351-0198, Japan.}

\altaffiltext{2}{Institute of Astronomy, Faculty of Science, The
  University of Tokyo, 2-21-1 Osawa, Mitaka, Tokyo 181-0015, Japan.}

\altaffiltext{3}{Princeton University Observatory, Peyton Hall,
  Princeton, NJ 08544.}  

\altaffiltext{4}{Institute of Space and
  Astronautical Science, Japan Aerospace Exploration Agency, 3-1-1
  Yoshinodai, Sagamihara, Kanagawa 229-8510, Japan.}

\altaffiltext{5}{Department of Physics and Astronomy, Johns Hopkins
  University, 3701 San Martin Drive, Baltimore, MD 21218.}

\altaffiltext{6}{Department of Physics, University of Tokyo, 7-3-1
  Hongo, Bunkyo, Tokyo 113-0033, Japan.} 
  
\altaffiltext{7}{Department of Astronomy and Astrophysics,
  Pennsylvania State University, 525 Davey Laboratory, University
  Park, PA 16802.}

\altaffiltext{8}{Institut d'Estudis Espacials de Catalunya/CSIC, Gran
  Capita 2-4, E-08034 Barcelona, Spain.}

\altaffiltext{9}{Department of Physics and Astronomy, York University,
  4700 Keele Street, Toronto, Ontario, M3J 1P3, Canada.}

\altaffiltext{10}{Department of Physics and Astronomy, Rutgers
  University, 136 Frelinghuysen Road, Piscataway, NJ 08854.  }

\altaffiltext{11}{Institute of Cosmology and Gravitation (ICG),
  University of Portsmouth, Portsmouth, PO1 2EG, UK.}

\altaffiltext{12}{Steward Observatory, University of Arizona, 933
  North Cherry Avenue, Tucson, AZ 85721. }

\altaffiltext{13}{Astronomy and Astrophysics Department, University of
  Chicago, 5640 South Ellis Avenue, Chicago, IL 60637.}

\altaffiltext{14}{Research Center for the Early Universe, School of Science,
 University of Tokyo, 7-3-1 Hongo, Bunkyo-ku, Tokyo 113-0033, Japan.}

\begin{abstract}
We present results from {\it Chandra} observations of SDSS~J1004+4112,
a strongly lensed quasar system with a maximum image separation of
$15\arcsec$. All four bright images of the quasar, as well as resolved
X-ray emission originating from the lensing cluster, are clearly
detected. The emission from the lensing cluster extends out to
approximately $1\farcm5$.  We measure the bolometric X-ray luminosity
and temperature of the lensing cluster to be $4.7 \times 10^{44}~{\rm
  erg\,s^{-1}}$ and 6.4~keV, consistent with the
luminosity-temperature relation for distant clusters.  The mass
estimated from the X-ray observation shows excellent agreement with
the mass derived from gravitational lensing.  The X-ray flux ratios of
the quasar images differ markedly from the optical flux ratios, and
the combined X-ray spectrum of the images possesses an unusually
strong Fe~K$\alpha$ emission line, both of which are indicative of
microlensing.
\end{abstract}


\section{Introduction}

The quadruply lensed quasar SDSS~J1004+4112, first identified as a
gravitational lens candidate from optical imaging and spectroscopy
from the Sloan Digital Sky Survey \citep[SDSS;][]{york00}, has an
exceptionally large image separation of $\sim15\arcsec$
\citep{inada03,oguri04}. The $z=1.734$ quasar is multiply imaged by a
cluster of galaxies (rather than a galaxy) at $z=0.68$. This is the
only known quasar strongly lensed by a central part of a massive
cluster. The quasar appears to be radio-quiet since it is not detected
in radio surveys such as the FIRST survey \citep{becker95}. Subsequent
observations revealed additional intriguing aspects of the system.
 \citet{richards04} discovered time variability in the blue wings
  of several optical broad emission lines, which can be plausiblly
  interpreted in terms of microlesning.  {\it Hubble Space Telescope}
images detected the fifth image \citep{inada05} and multiply imaged
background galaxies \citep{sharon05}. SDSS~J1004+4112 appears to be an
ideal laboratory for exploration of the structure of a large cluster
of galaxies and a quasar.

The uniqueness of SDSS~J1004+4112 affords advantages that strongly
argue for multiwavelength observations, of which X-ray observations
constitute an essential part. Measurement of the X-ray properties of
the lensing cluster offers an opportunity to improve lens models of
SDSS~J1004+4112. Moreover, the lensing cluster is the first example of
a {\it strong-lens selected} cluster; it is interesting to determine
if the cluster follows the empirical scaling relations between
luminosities, temperatures, and masses. Second, we can investigate the
magnification ratios of the quasar images at X-ray wavelengths, which
may aid the interpretation of the microlensing event that has been
observed at optical wavelengths \citep{richards04}. Indeed, X-ray flux
ratios of lensed quasars are frequently different from optical flux
ratios; this phenomenon has been attributed to microlensing
\citep{chartas04,blackburne06}.

In this paper, we report our analysis of an observation of SDSS~
J1004+4112 with the {\it Chandra X-Ray Observatory}.  A key feature of
{\it Chandra} images is the high angular resolution (on the order of
$1\arcsec$); this resolution is required to allow reliable separation
of the X-ray flux of the lensing cluster from that of the quadruply
imaged quasar.  In this paper we will adopt a cosmological model with
the matter density $\Omega_M=0.27$, the cosmological constant
$\Omega_\Lambda=0.73$, and the Hubble constant $H_0=70~{\rm
  km\,s^{-1}Mpc^{-1}}$ \citep{dns03}. At the redshift of the cluster
($z=0.68$), $1\arcsec$ corresponds to 7.16~kpc.  Unless otherwise
specified, quoted errors indicate the 90\% confidence range.

\section{Observations}

SDSS~J1004+4112 was observed for 80~ks with the {\it Chandra} Advanced
CCD Imaging Spectrometer \citep[ACIS;][]{gpg03} on 2005 January 1 and
2. The data were obtained with the ACIS S3 CCD operating in VFAINT
mode. This CCD has a $1024\times1024$ pixel format with an image scale
of $0\farcs492$~pixel$^{-1}$.  The target was offset from the nominal
aim point with a Y-offset of $-1\arcmin$, however, this has little
effect on the spatial resolution.  The CCD temperature during the
observations was $-120{\rm ^{\circ}C}$.

The data were processed using the standard software packages {\tt CIAO
  3.2.1} and {\tt CALDB 3.0.3}. In the analysis of diffuse emission
from the cluster (\S\ref{sec:cluster}), periods of high background
rates (defined by $>20$\% higher than the quiescent rates in the
2.5--7~keV band) were removed; the net exposure time is 62042~s. On
the other hand, the data without lightcurve-filtering (79987~s) are
used in the analysis of quasar components (\S\ref{sec:quasar}) since
the background counting rate is negligibly small for point sources.

\begin{figure}
\epsscale{.8}
\plotone{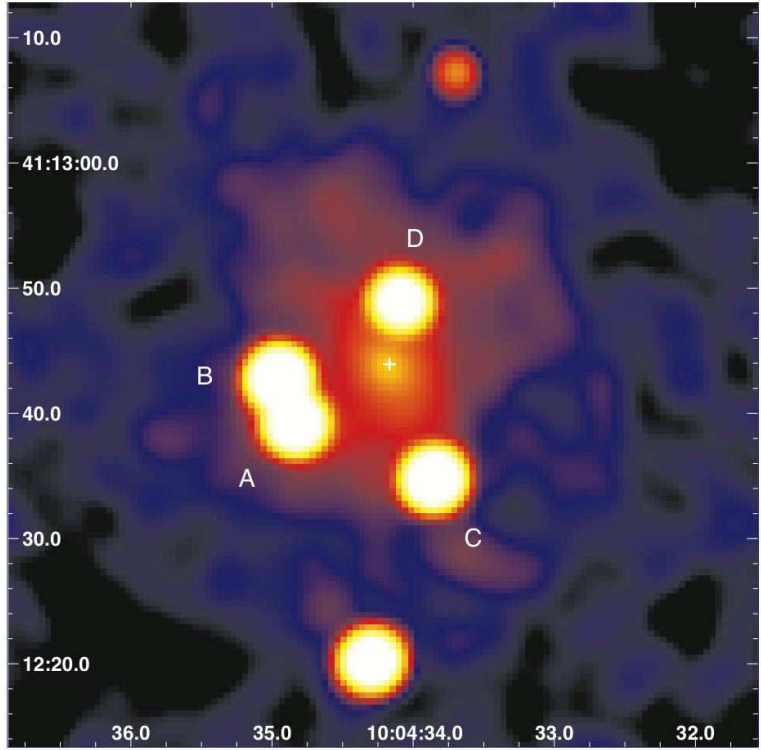}
\caption{Adaptively-smoothed ACIS-S3 image of SDSS~J1004+4112 in the
  0.5--7~keV band. Both multiple images of SDSS~J1004+4112, A--D, and
  the extended emission from the lensing cluster (whose X-ray peak is
  marked with the cross) are clearly seen. The images A and B, which
  have a minimum angular separation of $3\farcs8$, are also resolved
  in the raw image.  The point source at (10:04:34.294, +41:12:20.22)
  is a quasar (see text).
\label{fig:image}}
\end{figure}

The full band (0.5--7~keV) ACIS image is shown in
Figure~\ref{fig:image}.  The four images of SDSS~J1004+4112 are
clearly detected and resolved.  We found astrometry offsets for
  the {\it Chandra} data, ($\Delta$RA, $\Delta$Dec)=($0\farcs4\pm0\farcs1$,
  $-0\farcs1\pm0\farcs1$), from a comparison of three sources that lie within
  $65\arcsec$ from the aimpoint with those identified by the SDSS
  survey.  After correcting for the mean offsets, the measured coordinates of
  the four X-ray quasar images are consistent with those obtained with
  {\it Subaru} \citep{oguri04} within $<0\farcs2$.  In addition,
extended emission, roughly centered on the quasar images and extending
out to $r\sim 1\farcm5$, is seen in the X-ray data. This component
should arise from hot gas in the lensing cluster.  The relative
offsets of the coordinates of image A and the peak of the diffuse
emission are ($\Delta$RA, $\Delta$Dec)$=$($-7\farcs4\pm1\farcs0$,
$4\farcs7\pm1\farcs0$); the location of the peak of the extended
emission is consistent with the position of the brightest cluster
galaxy G1.

In the ACIS-S3 field, 39 point-like sources including the four quasar
images were detected with the {\tt wavdetect} algorithm with a
significance threshold parameter of $10^{-6}$.  The bright source at
(10:04:34.262, +41:12:20.33) with an X-ray flux of $\sim
1.0\times10^{-13}~{\rm erg\,s^{-1}cm^{-2}}$ (0.5--7~keV) has been
observed serendipitously as part of the follow-up program to
spectroscopically monitor components A and B at APO (Richards et
al. 2004); a low S/N spectrum indicates a redshift of $z\sim1.26$,
which is marginally consistent with the second most likely photometric
redshift ($z=1.325$ with 26\% confidence). The power-law slope,
$\alpha_{\rm ox}$, connecting the rest frame 2500\AA~and 2~keV flux
densities, is $\sim -1.2$, which is roughly consistent with the
$\alpha_{\rm ox}$-UV luminosity relation of AGNs \citep{strateva05,steffen06}.
The source at (10:04:33.653, +41:13:07.24) has an X-ray counting rate
of $(4.4\pm0.7)\times10^{-4}~{\rm counts\,s^{-1}}$, and it is not a
counter lens image of the southern bright source as the separation
angle is too large.

\section{Lensing Cluster}\label{sec:cluster}

\subsection{Spectral Analysis}

We derive the spectrum of the cluster component by extracting the data
from the circular region within a radius of $1\arcmin$ that is
centered on the position of the galaxy G1. The fluxes from the four
quasar images and several additional point sources were removed from
the spectral integration area by excluding all regions within a radius
10 times the size of the point spread function (PSF) at the source
positions. Here the size of PSF is defined as the 40\%
encircled-energy radius at 1.5 keV at the source position.  The
background was estimated from the data in a surrounding annulus
($2\arcmin<r<2\farcm5$). The source and background counts within
$r<1\arcmin$ are $1629\pm54$ and $906\pm20$ respectively.  The
0.5--7~keV extracted spectrum is fitted with the MEKAL thin-thermal
plasma model \citep{mewe85,mewe86,kaastra92,liedahl95} utilizing the
{\tt XSPEC} version~11. In this exercise, the Galactic hydrogen column
density is fixed to $N_{\rm H}=1.13\times 10^{20}{\rm cm^{-2}}$
\citep{dickey90}.  The X-ray temperature is constrained to be
$kT=6.4^{+2.3}_{-1.4}$~keV and the metal abundance to be
$Z=0.21Z_\odot$ (the 90\% upper limit is $0.62Z_\odot$). The Galactic
absorption-corrected, 0.5--7~keV flux is $1.62\times10^{-13}~{\rm
  erg\,s^{-1}cm^{-2}}$ ($r<1\arcmin$).  The bolometric X-ray
luminosity within $r_{500}$ is estimated as $L_{\rm
  X}=4.7\times10^{44}~{\rm erg\,s^{-1}}$, where $r_{500}$ is defined
as the radius within which the average matter density is equal to
$\Delta_c=500$ times the critical density of the Universe at the
cluster redshift.  The $\beta$-model analysis in
\S~\ref{subsec:xray_sb} yields $r_{500}=0.79$~Mpc.  This luminosity is
lower than the mean value expected from the luminosity-temperature
relation of distant clusters, $L_{\rm
  X}=1.9^{+2.4}_{-1.2}\times10^{45}~{\rm erg\,s^{-1}}$ \citep{ota06},
but our estimated value is within the observed scatter of the data.
The reduced chi-square of the best-fit model is $\chi^2/{\rm
  dof}=31.4/42$.

To investigate the radial temperature profile, we further analyzed
spectra integrated from the inner $r<0\farcm23(\sim 100~{\rm kpc})$
region and the outer $0\farcm23<r<1\arcmin$ region of the lensing
cluster. The fitted temperatures are $kT=6.3^{+2.5}_{-1.7}$~keV and
$5.9^{+3.4}_{-1.5}$~keV, respectively. These results suggest that
there is not a strong temperature drop towards the cluster center,
indicating radiative cooling is not important in this cluster.

\subsection{X-ray Surface Brightness Profile}\label{subsec:xray_sb}

\begin{figure}
\plottwo{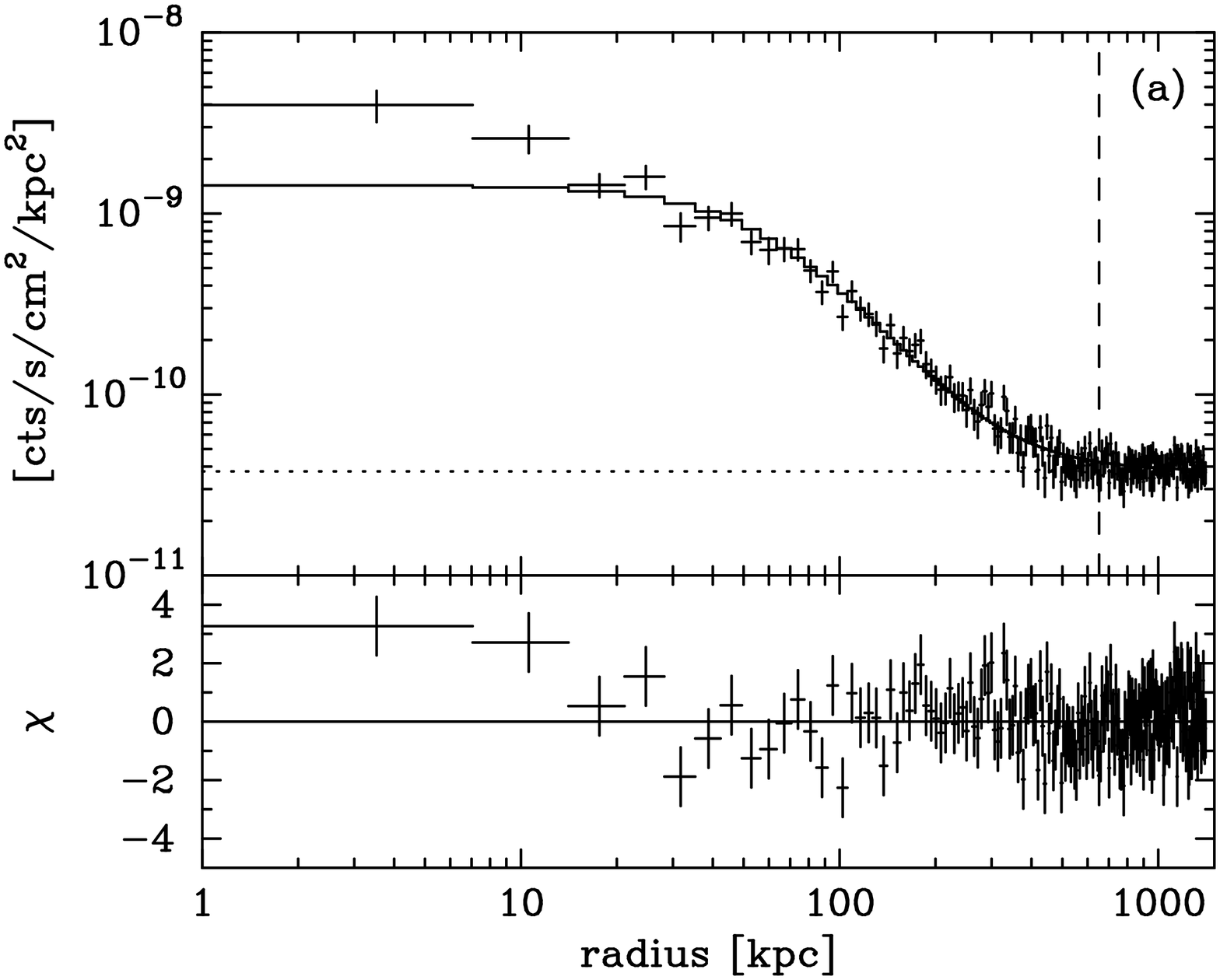}{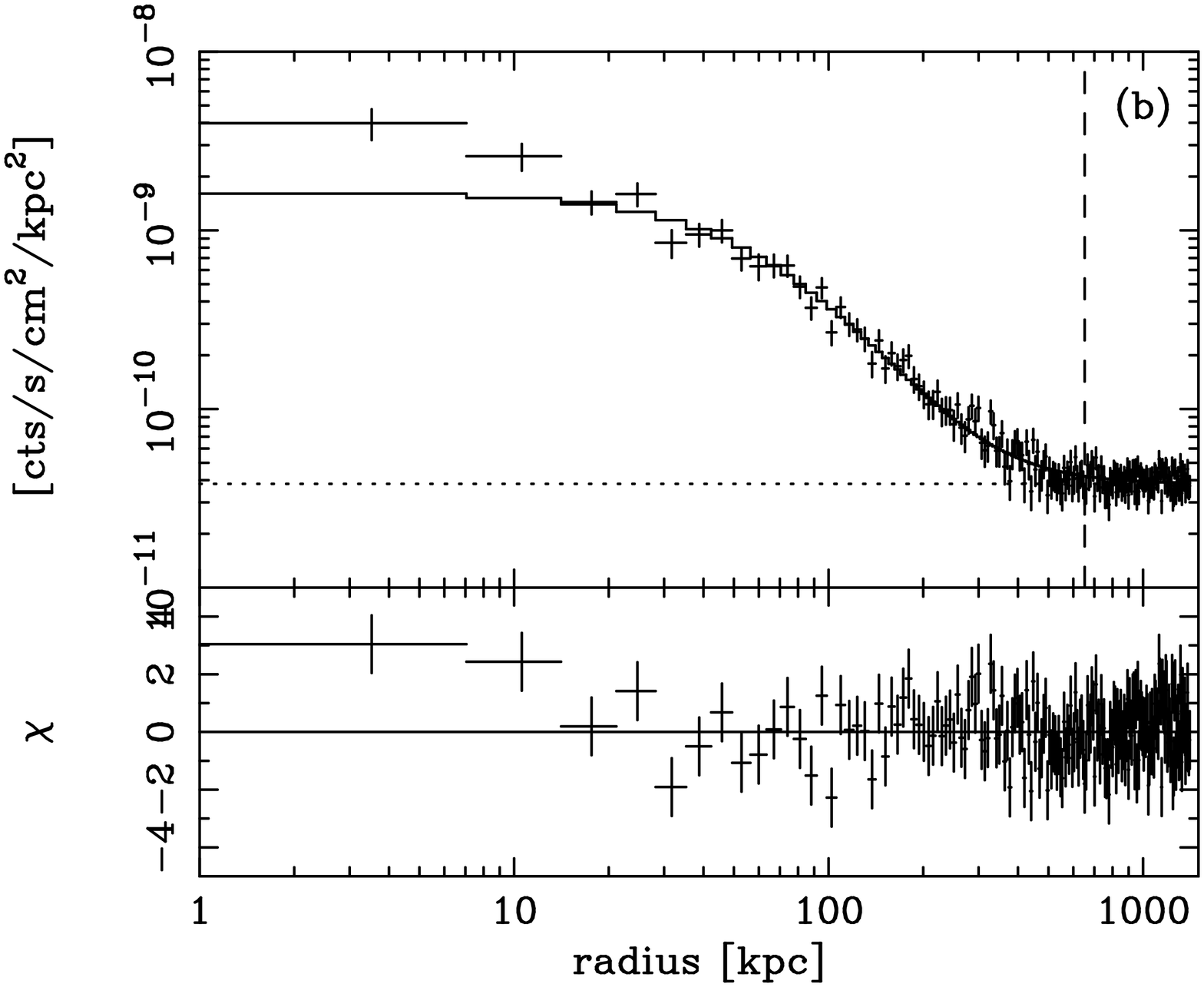}
\caption{Results of the X-ray surface brightness profile fitting with
  (a) the $\beta$-model and (b) the NFW-SSM model. In each panel, the
  crosses show the observed surface brightness in the 0.5--5~keV band
  and the solid line shows the best-fit model. The background is shown
  with the horizontal dotted line. The vertical dashed line shows the
  extent of the diffuse X-ray emission, $r_{\rm X}$ (see text).
\label{sb}}
\end{figure}

The one-dimensional radial surface-brightness profile of the extended
X-ray emission was constructed by adopting a center at G1 and
azimuthally averaging the 0.5--5~keV image, which was corrected for
the telescope's vignetting and the detector responses and rebinned by
a factor of two (i.e., 1 processed image pixel is $0\farcs98$).  
As shown in Figure~\ref{sb}, the radial profile was fit with the
following two models: (1) the conventional $\beta$-model
$S(r)=S_0[1+(r/r_c)^2]^{-3\beta+1/2}$, and (2) the profile derived
from the universal mass profile proposed by \cite{navarro97} plus
isothermality of the cluster \citep[][ hereafter NFW-SSM]{suto98}. The
background was set to a constant in the fitting.  We find that both
models can fit the observed radial profile reasonably well: The
reduced chi-square is $\chi^2/{\rm dof}=213.2/196$ and $205.5/196$ for
the $\beta$-model and NFW-SSM model, respectively. The fitted values
of the $\beta$-model parameters are $\beta=0.59^{+0.05}_{-0.04}$ and
$r_c = 9.7^{+2.0}_{-1.7}~{\rm arcsec} (= 69^{+15}_{-13}$~kpc), while
the NFW-SSM model results in $B=9.1^{+1.2}_{-0.8}$ and
$r_s=39^{+12}_{-9}~{\rm arcsec}(=278^{+85}_{-62}$~kpc). The profile of
the intra-cluster medium (ICM) has a compact core, typical of observed
surface brightness profiles of relaxed clusters.
 
The fit shows excess flux in the measurements compared to the models
within $r\lesssim 15$~kpc (Figure~\ref{sb}). This is often seen in
central regions of relaxed clusters \citep[e.g.,][]{mohr99}.  The
central emission is significantly extended compared with the PSF.
The 3-$\sigma$ upper limit on a point-source luminosity is
  $\sim4.4\times10^{42}~{\rm erg\,s^{-1}}$ (0.5--5~keV) assuming a
  power-law spectrum with index $\Gamma=2$ and a source redshift of
  $z=0.68$.  The excess luminosity within $r<2\arcsec$ is estimated
as $\sim 7\times10^{42}~{\rm erg\,s^{-1}}$ (0.3--8~keV), which is
higher than for a typical elliptical galaxy and rather comparable to
that of a cD galaxy in nearby clusters
\citep[e.g.,][]{matsushita00,osullivan01}. Note that we subtracted the
best-fit $\beta$-model component from the total emission assuming the
MEKAL model with $kT=6.4$~keV and $Z=0.21Z_\odot$ for the cluster
spectrum. The $B$-band luminosity of the galaxy G1 is $L_B\sim
3\times10^{11}L_\odot$ from the $g$ magnitude of 22.11 \citep{inada03}
and the color-transformation law for elliptical galaxies
\citep{fukugita95}. The above values may follow the $L_X-L_B$ relation
for elliptical galaxies \citep[e.g.,][]{osullivan01} and the excess
flux may be attributed to hot ISM emission from the central
galaxy. However, the X-ray temperature inferred from the hardness
ratio is $2.3^{+2.2}_{-1.0}$~keV assuming a $Z=1Z_{\odot}$ MEKAL
model. This is marginally high for a typical elliptical galaxy. Thus
we might be looking at a superposition of the galaxy component and
central emission from the cluster. Given the present photon
statistics, however, it is not possible to constrain further the
origin of the emission. Note that the total X-ray luminosity of
unresolved low-mass X-ray binaries is expected to be of the order
$\sim10^{41}~{\rm erg\,s^{-1}}$ \citep[e.g.,][]{kim04} and contributes
only $\sim 1$\% to the excess luminosity.  X-ray emission from the
central fifth image \citep{inada05} is unlikely: the contribution is
estimated to be only $\sim3-9$ counts (i.e., about $5-15$\% of the
central $r<2\arcsec$ emission) assuming that the X-ray emission from
the fifth image has similar intensity ratios as those measured in the
HST ACS image. The observed profile may be better fitted by
introducing the two-component $\beta$-models or increasing the inner
slope parameter $\alpha$ in the NFW-SSM model; however, we will not
pursue this process further since the above two models already
provided acceptable fits to the data. A considerably deeper
observation is needed to constrain further the gas profile in the
innermost region.

From the image analysis, the extent of the X-ray emission above the
3-$\sigma$ background level is found to be $r_{\rm
  X}=91\arcsec(=652~{\rm kpc})$ (Figure~\ref{sb}), which is
close to an overdensity radius of $r_{500}=0.79^{+0.15}_{-0.11}$~Mpc.

\subsection{Cluster Mass Distribution}
\begin{figure}
\epsscale{.8}
\plotone{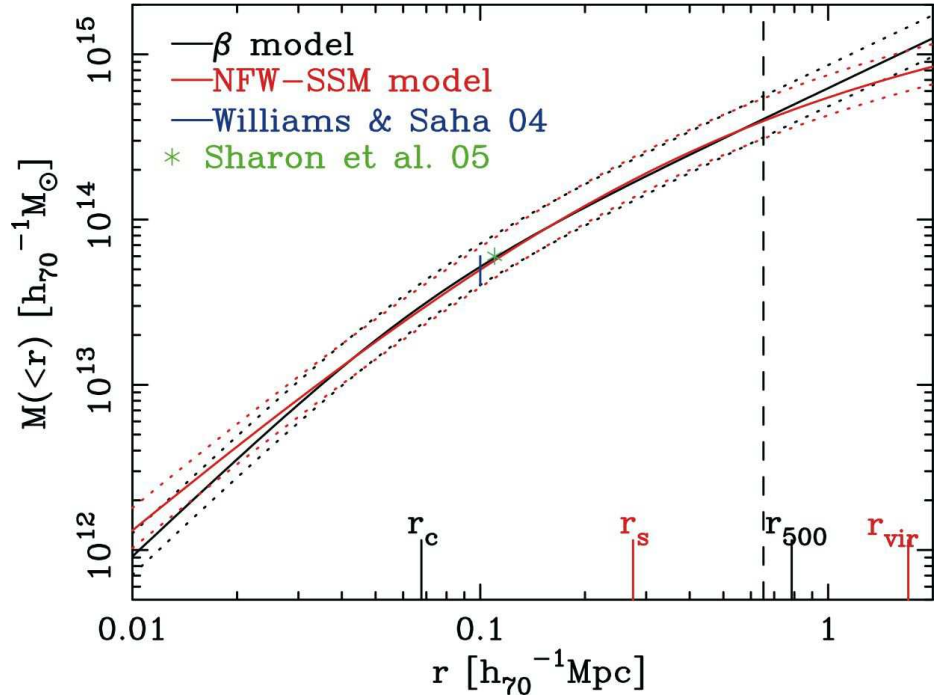}
\caption{Enclosed mass of the lensing cluster, $M_{\rm X}$, for the
  $\beta$-model (black) and the NFW-SSM model (red).  The dotted lines
  indicate the 90\% error ranges.  Note that $M _{\rm X}$ is a
  cylindrical cluster mass projected within a radius $r$. The masses
  derived from gravitational lensing \citep{willimas04, sharon05} are
  also shown for comparison. The meaning of the vertical dashed line
  is the same as Figure~\ref{sb}.
\label{fig:mx}}
\end{figure}

Under the assumption of hydrostatic equilibrium, we can infer the mass
distribution of the lensing cluster from the X-ray temperature and
surface-brightness profiles.  Because there is not a significant
radial dependence of the X-ray temperature in the observations, we
assume isothermality of the gas in the mass estimation.
Figure~\ref{fig:mx} shows the cylindrical cluster mass projected
within a radius $r$ derived from each surface mass distribution
profile. We find that both models yield consistent mass profiles
within the measurement errors.

It is interesting to compare the mass derived from X-ray analysis with
that from gravitational lensing. \citet{willimas04} derived the
cylindrical mass to be $M(<100~{\rm
  kpc})=(5\pm1)\times10^{13}M_\odot$, and \citet{sharon05} estimated
$M(<110~{\rm kpc})=6\times10^{13}M_\odot$. From Figure~\ref{fig:mx},
it is clear that both mass estimations are in excellent agreement with
the mass profile from the X-ray observation. For instance, we find the
cylindrical masses within 100~kpc for the $\beta$-model and NFW-SSM
model are $5.2^{+2.0}_{-1.2}\times 10^{13}M_\odot$ and
$5.0^{+1.8}_{-1.1}\times 10^{13}M_\odot$, respectively. The
discrepancy between X-ray and lensing masses has been reported in many
lensing clusters \citep[e.g.,][]{hattori99}, and is often ascribed to
the projection of the extra matter along line-of-sight, the elongation
of the lensing cluster along the line-of-sight direction, or a
departure from equilibrium. The excellent agreement between X-ray and
lensing masses implies that none of these effects is significant for
SDSS~J1004+4112.

For the NFW-SSM model and $\Delta_c=18\pi^2\Omega^{0.427}$
\citep{nakamura97}, the virial mass and the concentration parameter
are constrained to be $M_{\rm vir}=6.0^{+3.7}_{-2.1}\times
10^{14}M_\odot$ and $c_{\rm vir}=6.1^{+1.5}_{-1.2}$. The value of the
concentration parameter is slightly larger than the theoretically
expected median value for this virial mass and redshift, $c_{\rm
  vir}\sim 4.0$ \citep{bullock01}, but is within 2-$\sigma$ scatter
among different clusters.

\section{Quasar Images}\label{sec:quasar}

\subsection{Anomalous Flux Ratios}\label{sec:xrayflux}

The observed source counts for the four images, A--D, are 1237, 1580,
1312, and 763, in the 0.5--7~keV band, respectively.  Light curves
were produced with time resolutions of 2048 or 4096 sec.  We did not
find any clear sign of time variability within the statistical
uncertainties.  Note that each component was extracted using a
circular region with a radius of $2\arcsec$ in the analysis presented
in this section. The sum of the background and contamination from the
cluster emission are estimated to be only $\sim1$\% of the source
counts for A--C, and $\sim3$\% for D, and are negligibly small
compared to the statistical errors of the source spectra.

We measure the energy fluxes of lensed images in the {\it Chandra}
data by fitting each spectrum with a power-law plus Gaussian line
model (see the next subsection for details). The results are
summarized in Table~\ref{tab:flux}. The X-ray flux ratios differ
significantly from those measured in the optical \citep{inada05}. 
Since for fold lenses such as SDSS~J1004+4112, the
two images near the critical curve are usually brighter than the other
two images and since the optical flux ratios were reproduced with
simple mass models \citep{oguri04}, we believe that the X-ray, rather
than the optical, ratios are ``anomalous''. The optical/X-ray flux
ratios of images C and D are almost the same, therefore the most
natural interpretation is that the image A is demagnified by a factor
of $\sim 3$ in X-rays.  This interpretation is consistent with the
observation that image A appears to be a saddle-point image
\citep{oguri04,willimas04} which is more likely to be demagnified by
perturbations than the other images \citep{schechter02}.  While
  the optical data compared in Table~\ref{tab:flux} were taken on 2004
  April 28 \citep{inada05}, the demagnification of image A in X-rays
  is further supported by optical observations on 2004 November 27 and
  2005 April 26 with the MAGNUM telescope \citep{yoshii02}. By
  linearly interpolating these two $R$-band imaging observations, we infer  
  optical flux ratios on 2005
  January 1 to be B/A=0.69, C/A=0.40, 
and D/A=0.29, which are not very different from those listed in Table~\ref{tab:flux}.

In Table~\ref{tab:flux}, we also show the values of $\alpha_{\rm ox}$
to check the consistency of the X-ray and optical fluxes. The values
are somewhat higher than the $\alpha=-1.33$ that is expected from the
UV luminosity of SDSS~J1004+4112, $l({\rm 2500\,\AA})=10^{29.0}~{\rm
    [erg\,s^{-1}\,Hz^{-1}]}$ \citep[assuming a magnification
  factor of $\mu=50$ for image A;][]{oguri04} and the best-fit
$\alpha_{\rm ox}$-UV luminosity relation \citep{steffen06}.  From
Table~5 of \cite{steffen06}, $\alpha_{\rm ox}$ is
$-1.408\pm 0.165$ and $-1.322\pm0.192$ for $\log{l({\rm 2500\,\AA})}=29-30$ 
and $\log{l({\rm 2500\,\AA})}=28-29$, respectively. Thus the
derived $\alpha_{\rm ox}$ is within the 1-$\sigma$ (2-$\sigma$)
scatter for A (B--D).

\subsection{X-ray Spectrum}\label{sec:xrayspec}

\begin{figure}
\epsscale{1.}
\plottwo{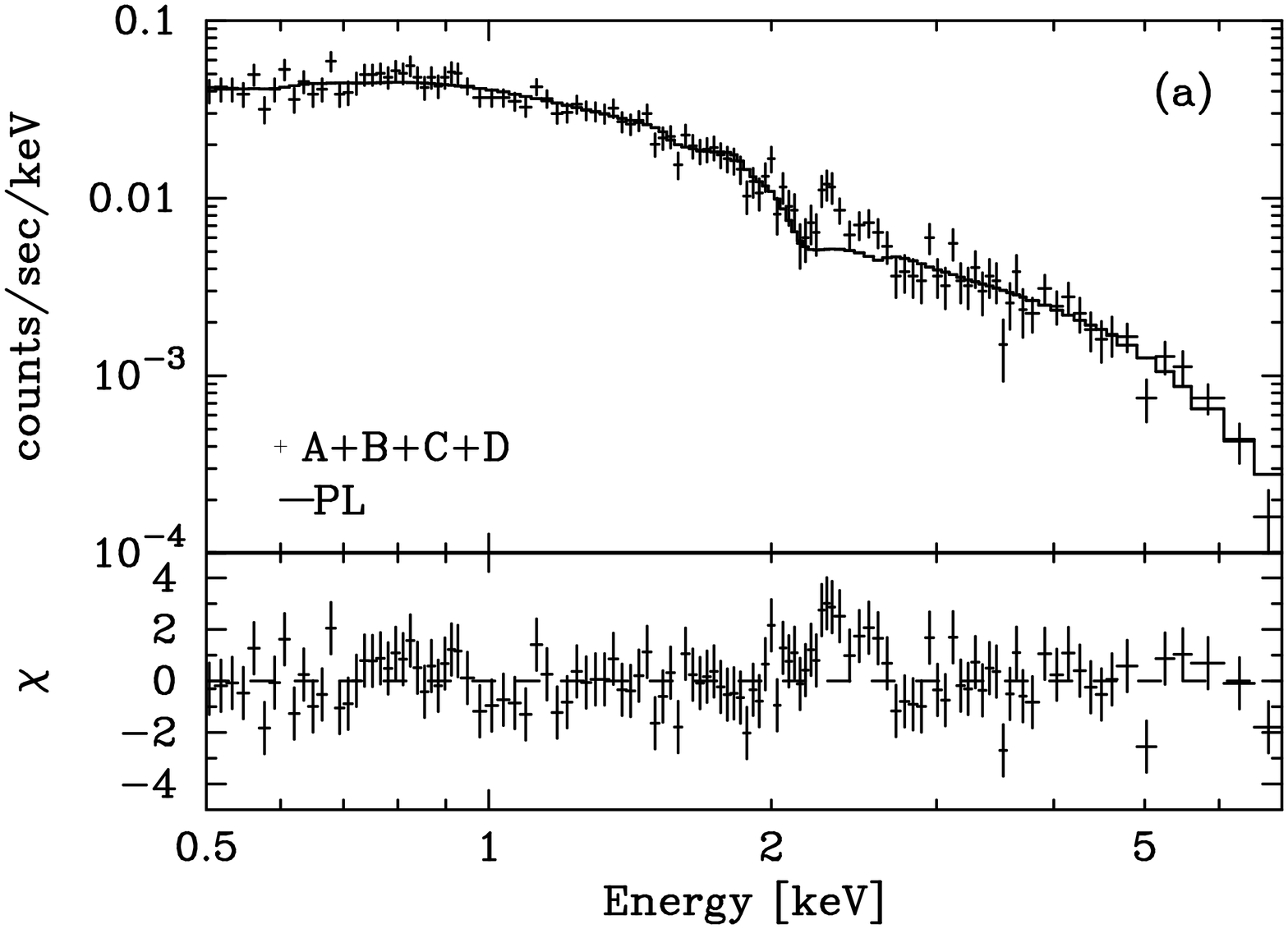}{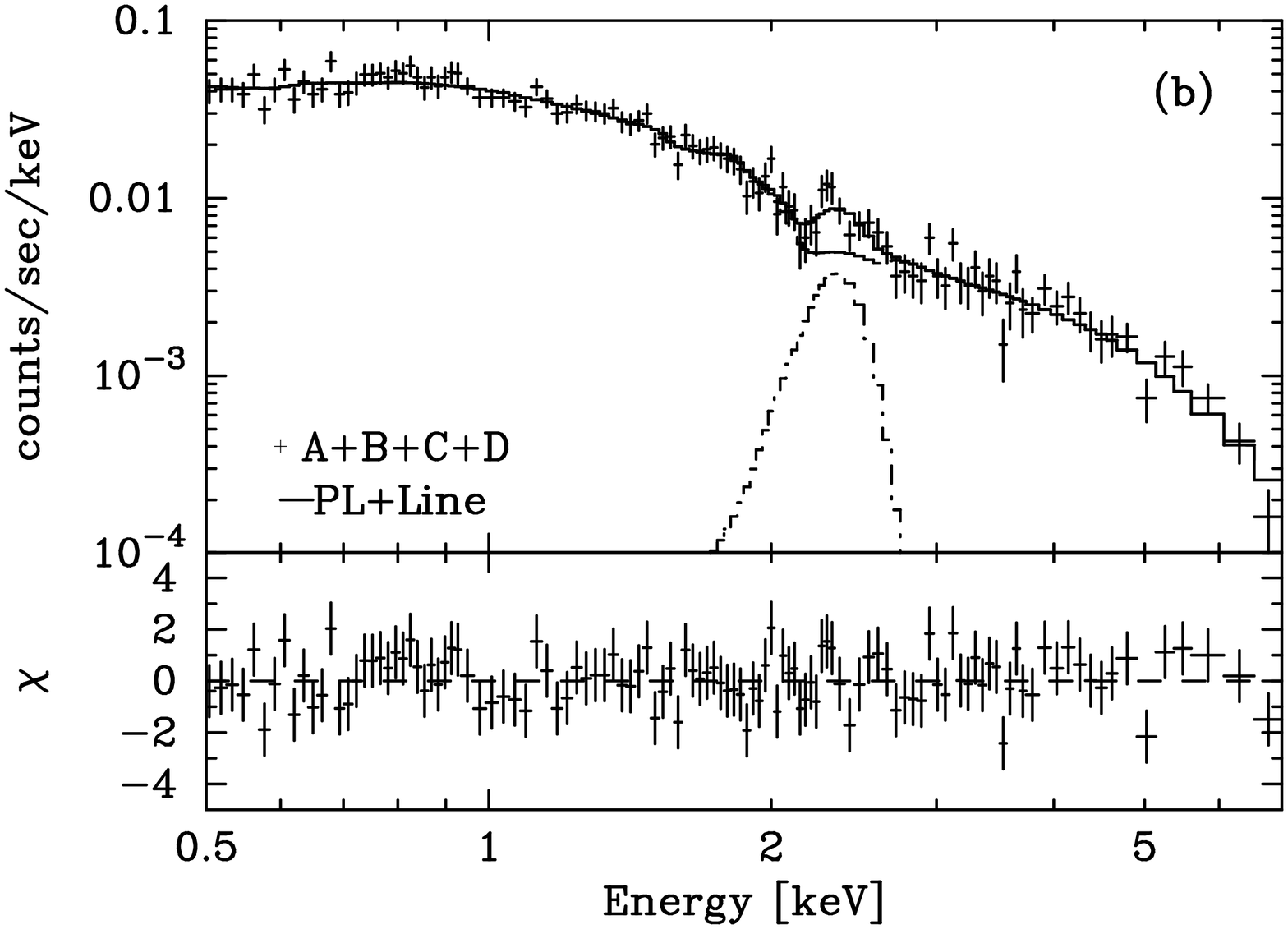}
\caption{Total quasar spectrum (the sum of images A, B, C, and D) fit  
with a power-law ({\it dashed}) plus Gaussian ({\it dot-dashed})
model, which is shown by the solid line.  
\label{fig:qso}}
\end{figure}

The spectrum of the lensed quasar component is displayed in
Figure~\ref{fig:qso}. To enhance the signal-to-noise ratio, the total
spectrum (sum of the images A, B, C, and D) is plotted.   We fit
  the spectrum with a power-law model. Again, the Galactic hydrogen
  column density is fixed to $N_{\rm H}=1.13\times 10^{20}{\rm
    cm^{-2}}$ since we did not find significant intrinsic
  absorption. This model is rejected at the 90\% confidence level
  ($\chi^2/{\rm dof}=144.0/115$). We plot the residual of the data against
  the best-fit model in Figure~\ref{sb}a, which shows 
  $\sim2-3\sigma$ excesses in adjacent bins around 2.3~keV in our
  frame. Then we mask these bins and those to either side (9 bins from 
  2.26--2.66 keV) and fit a power-law model. The chi-square becomes
  significantly smaller ($\chi^2/{\rm dof}= 99.5/106$).  The above
  results suggest the presence of an emission line component at
  $\sim2.3$~keV.  We then fit the spectrum with a power-law plus
Gaussian line profile model (see Figure~\ref{fig:qso}b).  In
comparison to just a power-law model, the fit was improved at the
$>99.99$\% confidence level according to the $F$-test
($\Delta\chi^2=35.9$ for three additional parameters).  The fitting
yields the power-law index $\Gamma=1.90\pm0.04$, the centroid energy
of the line $E=6.32^{+0.15}_{-0.14}$~keV, and the Gaussian width of
the line $\sigma=355^{+128}_{-169}$~eV.  The line centroid is
consistent with the neutral iron K$\alpha$ line at 6.4~keV in the
quasar rest frame and the derived $\sigma$ suggests that the line is
intrinsically broad. From the result, we can compute the equivalent
width ${\rm EW}=768^{+236}_{-227}$~eV in the quasar rest frame.  The
luminosity is $4.8\times10^{45}~{\rm erg\,s^{-1}}$ in the 2--10~keV
band.  In the case of $\mu=50$ for the image A, though the lens
  models that are consistent with the data allow a wide range of $\mu$
  \citep{oguri04}, the intrinsic quasar luminosity is estimated to be
  $\sim2.4\times10^{43}~{\rm erg\,s^{-1}}$ (2--10~keV). This is within the 
  luminosity range of Seyfert galaxies \citep[e.g.,][]{green92}. The
reduced chi-square of the best fit model is $\chi^2/{\rm
  dof}=108.1/112$.

There is a well-known negative correlation between the strength of the
iron line and the X-ray luminosity in quasars
\citep[e.g.,][]{nandra97}.  The derived equivalent width is larger by
a factor of $\sim 3$ than that expected from the relation of
\cite{nandra97} and a factor of $\gtrsim4$ than that from recent {\it
  XMM-Newton} observations \citep{page04, jimenez-bailon05}. Combined
with the result in \S \ref{sec:xrayflux}, we speculate that
microlensing demagnifies the X-ray continuum of image A, while keeping
(or amplifying) the Fe K$\alpha$ line.  Moreover, microlensing of
emission lines in the optical band has been detected for this lens
system \citep{richards04}, providing additional evidence for the
microlensing hypothesis.
The relative enhancement of Fe K$\alpha$ line has often been
observed in the X-ray emission of lensed quasars
\citep{oshima01,chartas02,chartas04,dai03}, and interpreted in terms
of microlensing. Indeed, the relative enhancement of Fe K$\alpha$ line
can be reproduced if the emission is separated in two regions
\citep[e.g.,][]{popovic06}.

\begin{figure}
\epsscale{0.4}
\plotone{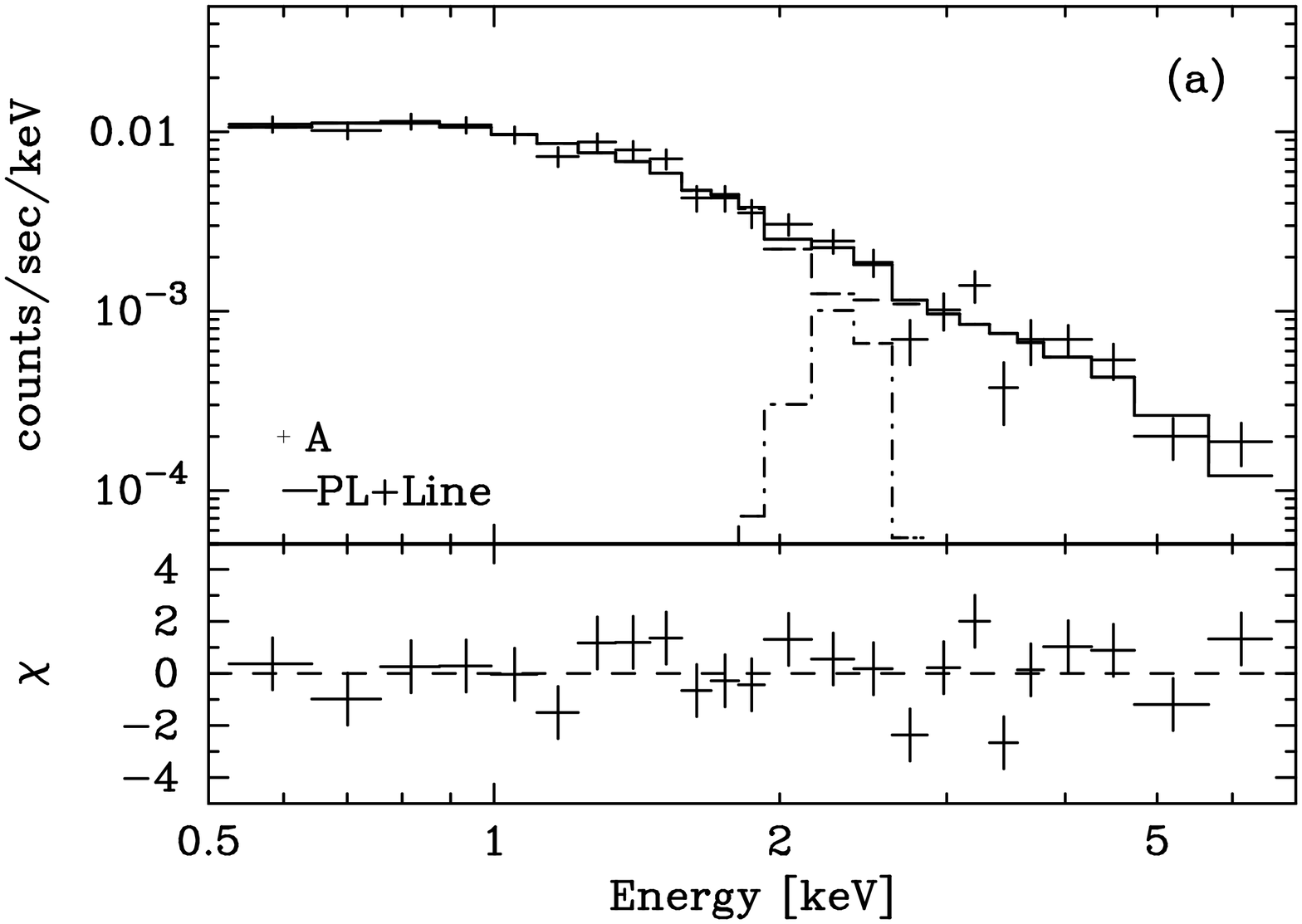}
\plotone{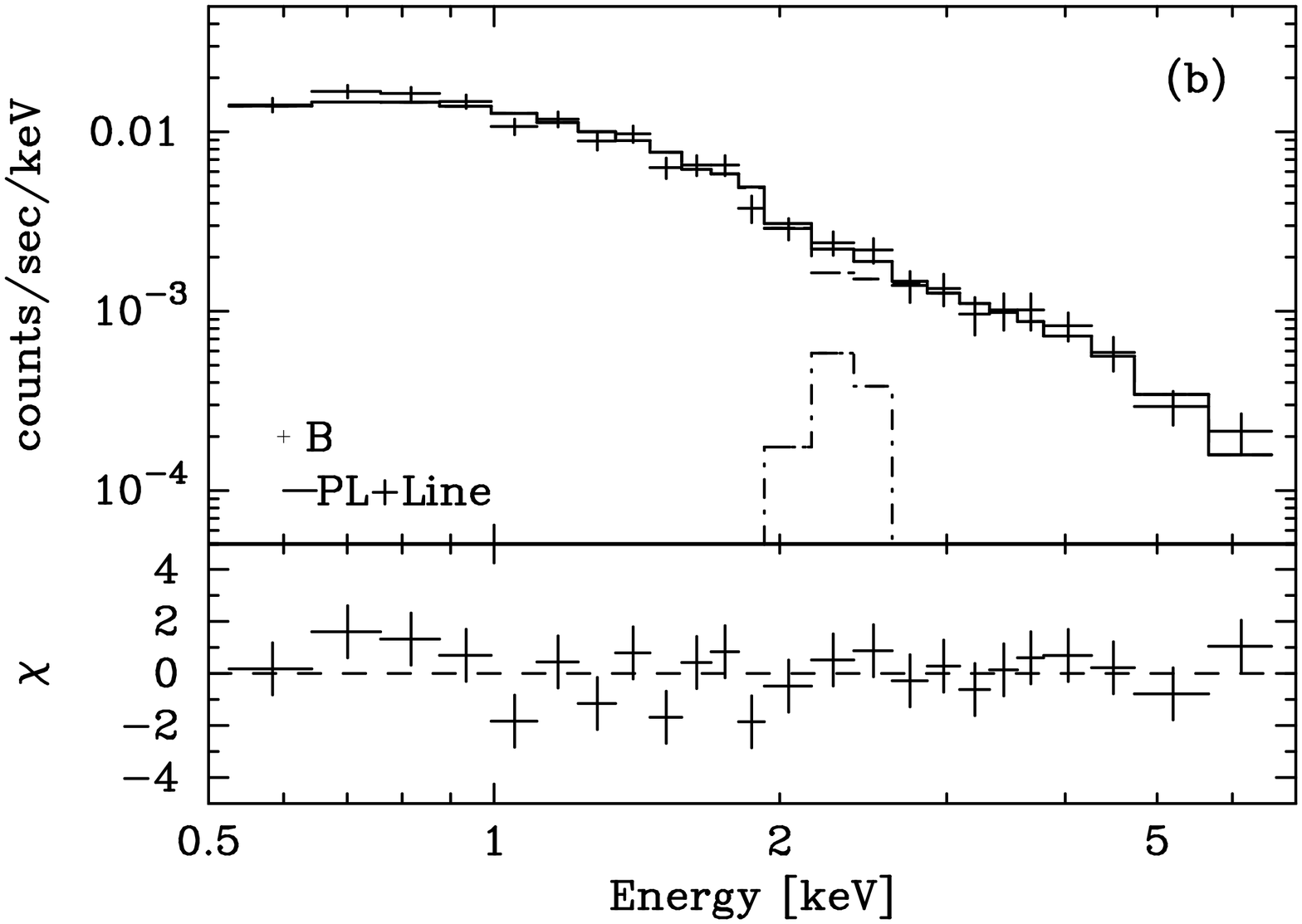}
\plotone{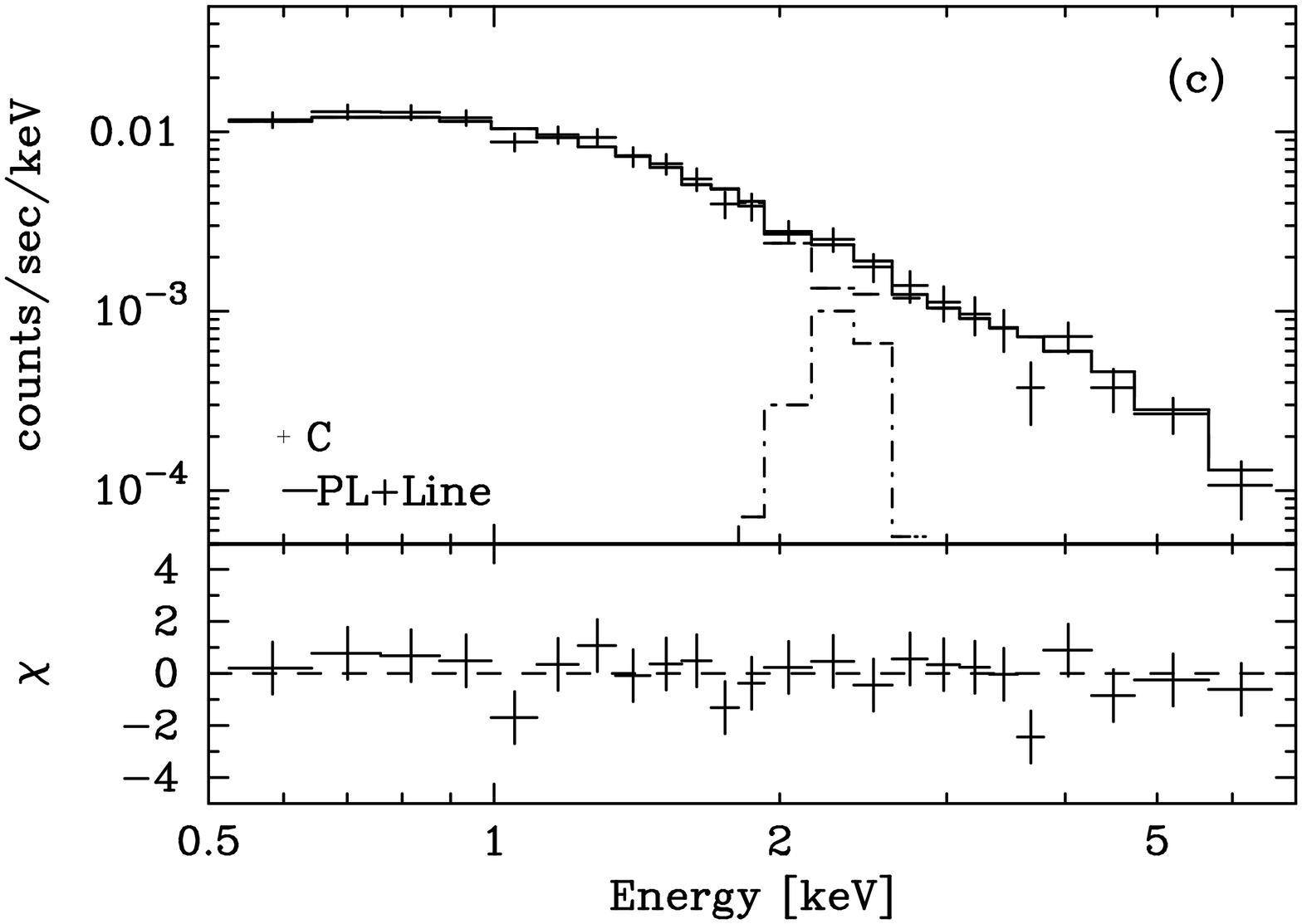}
\plotone{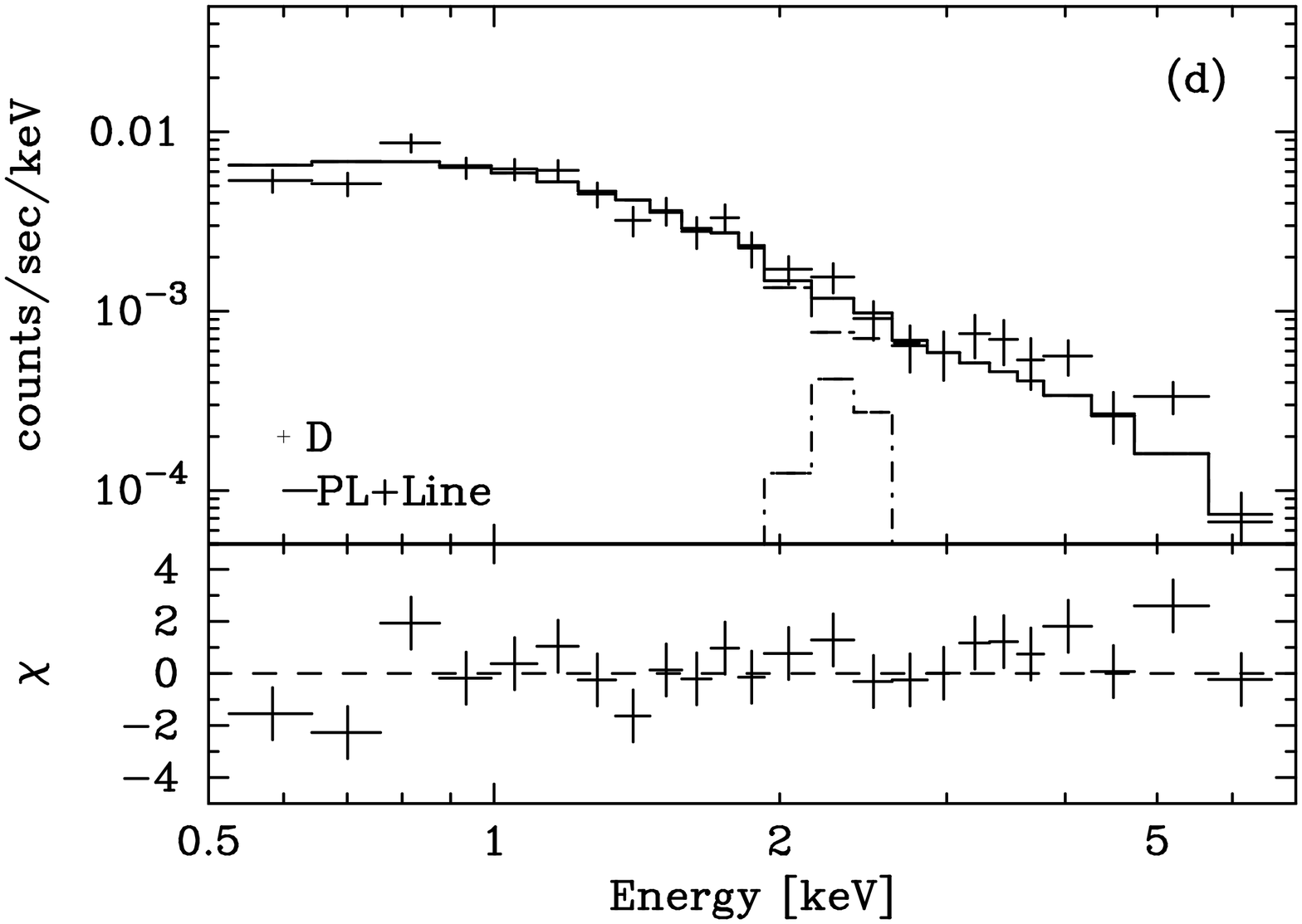}
\caption{X-ray spectra of images A -- D fit with a power-law ({\it
    dashed}) plus Gaussian ({\it dot-dashed}) model, which is shown by
  the solid line.  \label{fig:4spec}}
\end{figure}

 The origin of the strong Fe K$\alpha$ line is further explored by
  analyzing the spectrum of each image, rather than the sum of all
  images.  To measure the line intensities, we fit them with the
  power-law plus Gaussian model.  The centroid energy and the width of
  the line as well as the power-law index are allowed to vary in the
  fit. The best-fit parameters and the 90\% errors are listed in
  Table~\ref{tab:specfit}.  We found that the spectral parameters for
  the four images agree with each other at the 95\% confidence level,
  although the parameters, particularly the line width, show large
  uncertainties.  The fitting results are also statistically
  consistent with those derived for the total spectrum. We then
  performed spectral fitting with $E$, $\sigma$ and $\Gamma$ fixed to
  the best-fit values for the total spectrum (Figure~\ref{fig:4spec})
  and calculated the equivalent widths (Table~\ref{tab:flux}). 
Although the errors are very large, the Fe K$\alpha$ line is detected
most significantly for image A, supporting the interpretation above.
We also find that the hardness ratios of the images are marginally
consistent with each other (see Table~\ref{tab:flux}).

 \cite{green06} recently proposed that spectral differences
  between lensed quasar image components are due to small
  line-of-sight differences through quasar disk wind outflows. In
  particular, the author suggested that SDSS~J1004+4112 components B-D
  suffer from absorption in a disk-wind outflow and that the blue
  enhancements in SDSS~J1004+4112 A \citep{richards04} could be due to
  the alleviation of absorption along that sightline.  Indeed, the A-B
  diffference spectrum from Keck/LRIS spectroscopy of C IV region
  \citep{oguri04} resembles that of the troughs seen in a broad
  absorption line (BAL) quasar.  However, the observed line profiles
  and transitions involved (e.g., He II) mean that these features more
  plausibly result from excess {\em emission} in component A rather
  than absorption in component B. In addition, quasars with intrinsic
  UV absorption (e.g., BAL troughs) are generally X-ray faint due to a
  large intrinsic X-ray absorption \citep[$N_{\rm H}\sim
  (0.1-10)\times10^{23}~{\rm cm^{-2}}$;e.g.,][]{brandt00,
    green01,gallagher02,gallagher06}, however, intrinsic absorption
  was not significantly detected in the X-ray spectra of the four
  images; the 90\% upper limit on the column density is obtained to be
  $N_{\rm H} < 5\times 10^{21}~{\rm cm^{-2}}, 5\times 10^{21}~{\rm
    cm^{-2}}, 4\times 10^{21}~{\rm cm^{-2}}$, and $8\times
  10^{21}~{\rm cm^{-2}}$ for A--D, respectively. Here we assumed
  neutral absorbing material at $z=1.734$. In the case that the
  absorber is highly ionized and has high velocity flow, as proposed
  by \cite{green06}, X-ray spectra with considerably better statistics
  are required to test the model.  Moreover, if the differential
  absorption is responsible for the different X-ray spectra of A and
  B, we expect (and \cite{green06} predicts) that the image A should
  be less absorbed in the X-ray. Instead, however, component A has the
  smallest X-ray to optical flux ratio and the X-ray flux ratio is
  ${\rm A/B}=0.79\pm0.03$ (the 1-$\sigma$ error; see
  Table~\ref{tab:flux}).  Thus, given multiple independent
  indications, we conclude that the microlensing phenomenon is the
  most natural interpretation for the present data.  

\section{Summary}

We have presented results from {\it Chandra} observations of SDSS
J1004+4112, a large-separation gravitational lens system created by a
cluster of galaxies.  We have detected X-ray emission from the lensing
cluster as well as four lensed quasar images.

From the cluster X-ray emission, we have constrained the bolometric
luminosity and the temperature to be $L_{\rm X}=4.7\times 10^{44}~{\rm
  erg\,s^{-1}}$ and $kT=6.4^{+2.3}_{-1.4}$~keV, consistent with the
luminosity-temperature relation of distant clusters. We have
reconstructed the mass profile of the lensing cluster assuming
isothermality and hydrostatic equilibrium, and found that the mass
within 100~kpc excellently agrees with that expected from strong
lensing.

X-ray emission from the lensed quasar images displays two anomalies:
the presence of a strong Fe K$\alpha$ line and significant differences
in the flux ratios from those found in the optical band. Both of these
features are suggestive of microlensing. The idea is supported by the
fact that image A, which appears most anomalous, is a saddle-point
image.

\acknowledgements We are grateful to A. Yonehara for useful
discussions. N.~O. acknowledges support from the Special Postdoctral
Researchers Program of RIKEN. N.~I. is supported by JSPS through JSPS
Research Fellowship for young scientists.


\begin{deluxetable}{ccccccc}
\tablewidth{0pt}
\tablecaption{X-ray and Optical Properties of SDSS~J1004+4112\label 
{tab:flux}}
\tablehead{\colhead{Name}
& \colhead{$F_{\rm X}$\tablenotemark{a}}
& \colhead{hardness\tablenotemark{b}}
& \colhead{$F_{\rm opt}$\tablenotemark{c}}
& \colhead{$F_{\rm opt}/F_{\rm X}$}
& \colhead{$\alpha_{\rm ox}$}
& \colhead{EW [eV]\tablenotemark{d}}}
\startdata
A & $0.95\pm0.03$ & $0.30\pm0.02$ & 1.00 & 1.00 & $-1.27\pm0.01$ & $1151^{+ 374}_{-554}$\\
B & $1.21\pm0.03$ & $0.26\pm0.02$ & 0.73 & 0.58 & $-1.16\pm0.01$ & $ 593^{+ 284}_{-446}$\\
C & $1.01\pm0.03$ & $0.27\pm0.02$ & 0.35 & 0.33 & $-1.12\pm0.01$ & $842^{+ 443}_{-389}$\\
D & $0.56\pm0.02$ & $0.34\pm0.03$ & 0.21 & 0.35 & $-1.10\pm0.01$ & $ 552^{+ 369}_{-268}$\\
\enddata
\tablenotetext{a}{Absorption-corrected, 0.5--7~keV X-ray flux in units of 
  $10^{-13}{\rm erg\, s^{-1}cm^{-2}}$ and the 1-$\sigma$ error.}
\tablenotetext{b}{Hardness ratios defined by the count rate ratios in
  $2-7$~keV and $0.5-2$~keV bands. Errors indicate 68\% confidence limits.}
\tablenotetext{c}{Flux normalized by the flux of image A in the HST
  F814W image \citep{inada05}. Measurement errors are negligibly
  small. }  
\tablenotetext{d}{Equivalent width of the Fe K$\alpha$ line in the
  quasar rest frame.}
\end{deluxetable}


\begin{deluxetable}{ccccccc}
\tablewidth{0pt}
\tablecaption{Results of spectral fitting for quasar components\label{tab:specfit}}
\tablehead{\colhead{Name}
& \colhead{$\Gamma $\tablenotemark{a}}
& \colhead{$E$ [keV]\tablenotemark{b}}
& \colhead{$\sigma$ [keV]\tablenotemark{c}}
& \colhead{EW [eV]\tablenotemark{d}}
& \colhead{$\chi^2/{\rm dof}$}
& \colhead{$\Delta\chi^2$}
}
\startdata
A & $1.92^{+ 0.10}_{-0.09}$ & $6.15^{+ 0.26}_{-0.20}$ & $0.36^{+ 0.17}_{-0.28}$ & $1252^{+   516}_{  -496}$ & $31.0 / 20$  & 18.5\\
B & $1.97\pm0.08$ & $6.45^{+ 0.15}_{-0.11}$ & $<0.63$ & $615^{+   321}_{  -321}$ & $22.3 / 20$ & 9.6\\
C & $1.97\pm0.09$ & $6.25^{+ 0.35}_{-0.23}$ & $ 0.37(<0.76)$ & $1077^{+   522}_{  -535}$ & $14.7 / 20$ & 14.9\\
D & $1.77\pm0.11$ & $6.12^{+ 0.31}_{-0.27}$ & $0.26 (<0.62)$ &  $ 834^{+   521}_{  -575}$ & $30.8 / 20$ & 7.3\\
A+B+C+D & $1.90\pm0.04$ & $6.32^{+ 0.15}_{-0.14}$ & $0.36^{+0.13}_{-0.17}$ &  $ 768^{+   236}_{  -227}$ & $108.1 / 112$ & 35.9\\
\enddata
\tablenotetext{a}{Power-law index.}
\tablenotetext{b}{Centroid energy in the quasar rest frame.}
\tablenotetext{c}{Line width in the quasar rest frame.}  
\tablenotetext{d}{Equivalent width of the Fe K$\alpha$ line in the quasar rest frame.}
\tablenotetext{e}{Chi-square and the degree of freedom of the fit.}
\tablenotetext{f}{Improvement of the chi-square in comparison to the power-law model.}
\end{deluxetable}


\end{document}